\documentclass[manuscript]{acmart}

\AtBeginDocument{%
  }

\setcopyright{acmcopyright}
\copyrightyear{2018}
\acmYear{2018}
\acmDOI{XXXXXXX.XXXXXXX}

\acmPrice{15.00}
\acmISBN{978-1-4503-XXXX-X/18/06}

\usepackage{amsmath,amsfonts,bm}

\def\eqref#1{equation~\ref{#1}}

\def\1{\bm{1}}

\DeclareMathAlphabet{\mathsfit}{\encodingdefault}{\sfdefault}{m}{sl}
\SetMathAlphabet{\mathsfit}{bold}{\encodingdefault}{\sfdefault}{bx}{n}

\newcommand{\name}{\textit{DeclutterCam}}

\usepackage{bm}
\usepackage{wrapfig}
\usepackage{makecell}
\usepackage{csquotes}
\usepackage{subfigure}

\begin{document}

\title{DeclutterCam: A Photographic Assistant System with Clutter Detection and Removal}


\author{Xiaoran Wu}
\affiliation{%
 \institution{Tsinghua University}
 \streetaddress{30 Shuangqing Rd.}
 \city{Beijing}
 \country{China}}
 
\author{Zihan Yan}
\affiliation{%
 \institution{Massachusetts Institute of Technology}
 \streetaddress{77 Massachusetts Ave}
 \city{Cambridge}
 \state{MA}
 \country{United States}}
 
\author{Xiang ‘Anthony’ Chen}
\affiliation{%
 \institution{University of California, Los Angeles}
 \streetaddress{420 Westwood Plaza}
 \city{Los Angeles}
 \state{CA}
 \country{United States}}

\renewcommand{\shortauthors}{Wu et al.}

\newcommand{\xac}[1]
{{\fontfamily{cmss}\selectfont \color{orange} $\leftarrow$ {\bf XAC}: #1 }}


\begin{abstract}
Photographs convey the stories of photographers to the audience. However, this story-telling aspect of photography is easily distracted by visual clutter. Informed by a pilot study, we identified the kinds of clutter that amateurs frequently include in their photos. We were thus inspired to develop \emph{DeclutterCam}, a photographic assistant system that incorporates novel user interactions and AI algorithms for photographic decluttering. Clutter elements are detected by an aesthetic quality evaluation algorithm and are highlighted so that users can interactively identify distracting elements. A GAN-based iterative clutter removal tool enables the users to test their photographic ideas in real-time. User studies with 32 photography beginners demonstrate that our system provides flexible interfaces, accurate algorithms, and immediate feedback that allow users to avoid clutter and explore more photographic ideas. Evaluations by photography experts show that users can take higher quality photos that better convey the intended story using our system. 
\end{abstract}



\begin{teaserfigure}
    \includegraphics[width=\linewidth]{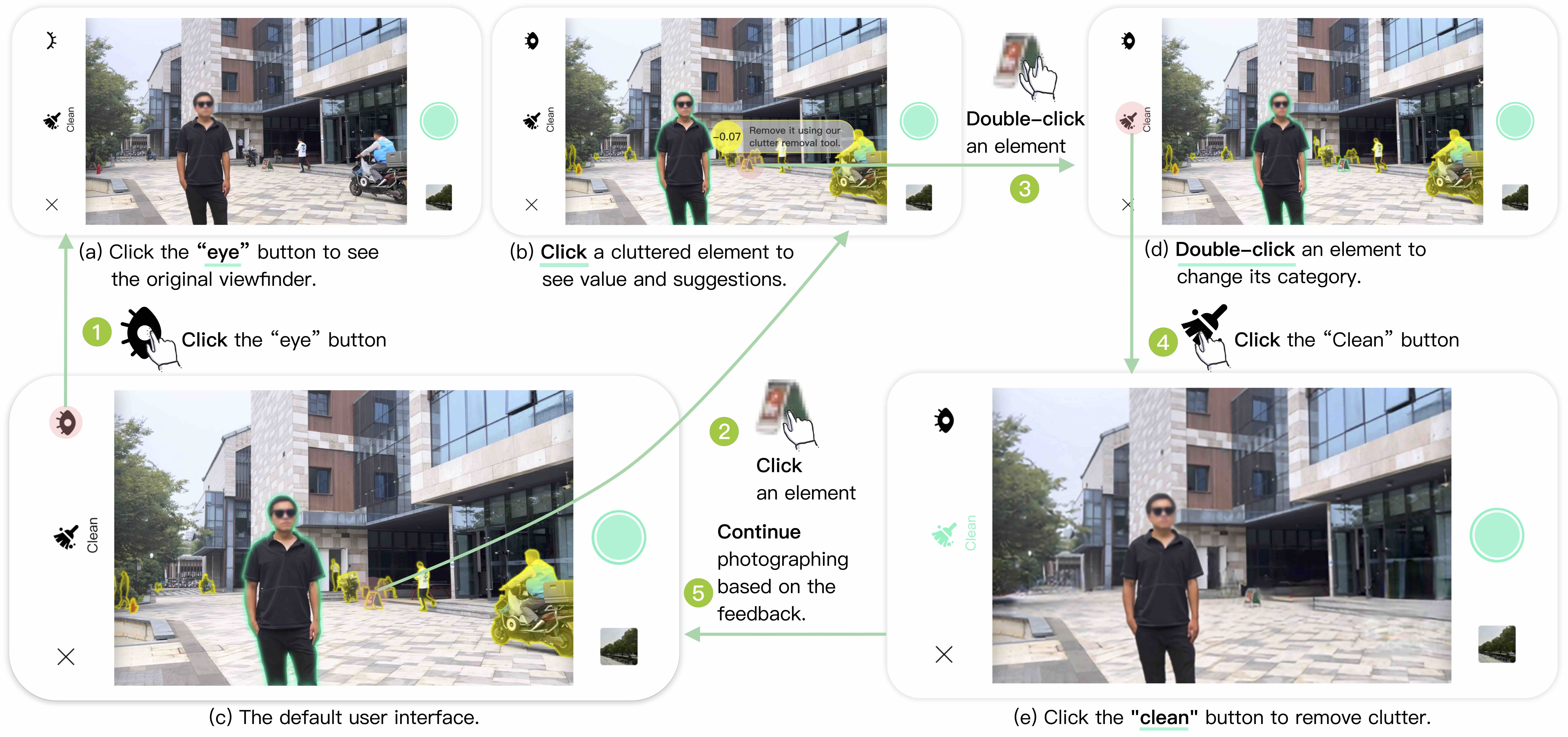}
    \caption{\name~is a photographic assistant system helping photography amateurs identify and handle clutter elements in photos. \textbf{(c)} On the default interface, detected clutter and other elements are highlighted by different masks. We develop a clutter identification model to support this interface. \textbf{(a)} By clicking the \emph{eye} button, users see the original image in the viewfinder. \textbf{(b)} The estimated contribution value to the overall aesthetic quality and content, as well as suggestions about how to deal with a cluttered element, can be obtained by clicking the element. \textbf{(d)} Users can change the category of an element by double-clicking it. \textbf{(e)} We develop an image inpainting algorithm that removes current clutter in the scene by clicking the "clean" button. With cleaned images as timely feedback, the user finds possible problems and explores more photographic ideas to take photos of higher quality.}
    \label{fig:interaction}
\end{teaserfigure}
\maketitle


\section{Introduction}

Photography can convey the emotions of photographers~\cite{kurtz2013happiness} to the audience by capturing a memorable moment, shooting a meaningful scene, or presenting an impressive portrait. This story-telling aspect~\cite{block2020visual,cohn2013visual,williams2019attending} of photographs features guiding feelings of an audience by giving visual clues~\cite{eisner2008graphic}, from which the observer discovers the larger context of photographs and perceives the intended feelings. However, amateur photographers often include distracting visual elements that prevent an audience from capturing intended clues and spoil the story of the photos.

With the increasing popularity of camera apps like those on mobile phones, users with limited photography knowledge or experience obtain easy access to photography. This trend leads to research interests in photographic guidance systems~\cite{mitarai2013interactive,wu2021tumera}. Among them, \citet{jane2021dynamic} focus on visual distractions in photos and build a system reminding users of objects in the photo by highlighting object edges. The decision of which objects are cluttered is left to users. However, it is unclear whether the provided interactions meet users' needs to identify and handle \emph{visual clutter}~\cite{xu2013structural}. Inexperienced users may struggle to evaluate the influence of visual elements on the story-telling aspect of photos and to figure out how to deal with them.


In this paper, we first carried out a pilot study to understand the types of visual clutter common to photography amateurs that affect story-telling and to investigate users' needs for the distinguishment and handling of these types of visual clutter. Inspiring results include that (1) photography beginners typically cannot identify visual elements that hurt story-telling, especially when they are familiar with the photo scene. (2) Even when identified, some cluttered elements are difficult to remove. Examples include numerous or immovable items in a scene, vehicles on a busy street, or endless tourists in a resort. 


This pilot study shows the gap between users' needs to better deal with clutter and the capability of current photography guidance systems. We are thus inspired to develop a photographic assistant system (\name) with capture-time clutter detection and automatic removal functions. The default user interface (Fig.~\ref{fig:interaction} (c)) is a viewfinder responding in real time. In this viewfinder, cluttered elements are distinguished from normal elements by masks with different colors and transparencies. To underpin this user interface, we develop a novel counterfactual computational model that quantifies the contribution of visual elements to the overall aesthetic quality and content of the photo.

When clicking a cluttered element (Fig.~\ref{fig:interaction} (b)), suggestions about how to remove it are given, including manipulating the camera view to circumvent clutters and using our AI-enabled clutter removal tool. Our clutter removal tool is activated on users' demand (Fig.~\ref{fig:interaction} (e)) for the clutter that cannot be easily removed. This tool generates preview images where the clutter is replaced with background textures. Users can use this preview image as the final photographic work or as feedback for their following photographic idea exploration. The clutter removal tool is powered by a novel iterative GAN-based image inpainting algorithm that progressively accepts generated image regions with high fidelity. The advantage of such a model is to generate realistic content for high-resolution images while fulfilling the real-time response requirement.


We validate \name~in a user study ($N=32$) with two levels of tasks: (1) In the module-level test, we focused on how users reacted to our clutter detection and removal functions. Peer reviews showed that participants found the clutter detection algorithm can distinguish clutter with a high success rate ($91.88\%$), and the clutter removal algorithm can generate visually realistic and semantically reasonable content. (2) In the system-level test, we evaluated whether using our system can help participants take photos of higher quality and better convey the intended stories. Peer and expert reviews, including quantitative rating and subjective feedback, demonstrate that our system can indeed reduce the clutter and help better convey the intentions of photographers. More importantly, participants reported that capture-time clutter detection and removal could encourage the photographic design process by providing timely, accurate, and helpful feedback. 
\section{Related Works}

Our work lies at the intersection of image processing, photography guidance, and image aesthetic evaluation. In this section, we discuss related work in these three fields of study.

\subsection{Capture-Time Photographic Guidance} In-camera guidance is a hot topic in human-computer interaction research. Previous work helps photographers by positioning their cameras for better compositions~\cite{bae2010computational,carter2010nudgecam,mitarai2013interactive,fried2020adaptive}, displaying view proposals~\cite{ma2019smarteye,rawat2015context}, guiding users to better lightning~\cite{jane2019optimizing,li2017guided}, and incorporating aesthetic evaluation models to provide suggestions on improving photos~\cite{wu2021tumera,wu2022interpretable}. Clutter is not the focus of these works.

Our work is closely related to \citet{jane2021dynamic}. In this paper, the authors also study the influence of clutter on photographs. They develop a system providing visual overlays highlighting the edges of objects. Beyond reminding the photographers of the existence of objects, we are interested in identifying and guiding the users to get rid of the clutter. Therefore, we introduce a computational evaluation model and define clutter as those objects contributing negatively to the aesthetic appearance and contents of the photo. We give suggestions about removing clutter. Furthermore, for those objects that can not be easily removed, we provide an image inpainting tool. These points make our system different from \cite{jane2021dynamic} for functions, interactions, visualization, and underlying techniques.

\subsection{Image Aesthetic Evaluation}
Our clutter detection module is related to previous work on computational aesthetics. Automatic image aesthetic assessments promote many real-world applications~\cite{westerman2007creative, wu2021tumera, wang2017deep, bhattacharya2010framework,datta2007learning} and human-computer interactions~\cite{tractinsky1997aesthetics,kurosu1995apparent, tullis1981evaluation,tullis1984predicting, kelster1983making,aspillaga1991screen,toh1998cognitive,szabo1999effects,heines1984screen,grabinger1993computer}. Conventional approaches rely on domain knowledge and designing \emph{handcrafted features} for evaluation. Hand-designed features generally belong to four categories. (1) Basic visual features, including color distribution~\cite{ke2006design,aydin2014automated}, hue count~\cite{ke2006design}, blurriness~\cite{tong2004classification}, sharpness~\cite{aydin2014automated}, depth~\cite{aydin2014automated}, global edge distribution~\cite{ke2006design}, low-level contrast~\cite{tong2004classification,ke2006design}, and brightness~\cite{ke2006design}. (2) Composition features~\cite{bhattacharya2010framework,dhar2011high,bhattacharya2011holistic,wu2010good,wu2010good,tang2013content,zhang2014fusion} that are designed to model techniques like the rule of thirds, low depth of field, and opposing colors. (3) General-purpose features that are not specifically designed for aesthetics assessment, such as Fisher vector (FV)~\cite{marchesotti2011assessing,marchesotti2013learning,murray2012ava}, and scale-invariant feature transform (SIFT)~\cite{yeh2012relative}. (4) Task-specific features that are designed specifically for a category of images. For example,~\citet{li2010aesthetic} and~\citet{lienhard2015low} design features for human faces, and \citet{su2011scenic} and~\citet{yin2012assessing} focus on landscape photos. Based on these features, heuristic rules~\cite{aydin2014automated}, SVMs~\cite{datta2006studying,luo2008photo,wu2010good,dhar2011high,wu2010good,yin2012assessing,tang2013content,lienhard2015low}, boosting~\cite{tong2004classification,su2011scenic}, regression~\cite{sun2009photo}, SVR~\cite{bhattacharya2010framework,li2010aesthetic,bhattacharya2011holistic}, naive Bayes classifiers~\cite{ke2006design}, and Gaussian mixture model~\cite{marchesotti2011assessing,marchesotti2013learning,murray2012ava} are used to get the overall aesthetic score estimation. For a detailed discussion of these methods, we refer readers to~\citet{deng2017image}. 

Handcrafted features make good use of human knowledge of aesthetics, but their design requires a large number of engineering efforts and is not flexibly applicable to different aesthetics tasks. Due to these shortcomings, they largely underperform methods based on automatic feature extraction by learning deep neural networks.

\citet{wang2016multi} modify AlexNet for \emph{deep aesthetic evaluation}. Specifically, the authors replace the fifth convolutional layer of AlexNet with a group of seven convolutional layers, each of which is dedicated to one category of scenes. \citet{tian2015query} train an aesthetic model with a smaller fully-connected layer as the feature extractor and an SVM as the classifier. Images are usually resized or cropped before being fed into the deep model because the model typically requires the input to be of the same dimensions. To prevent visual features from being corrupted during this process,~\citet{lu2015deep} randomly select image patches of the same size. These patches are directly fed into the deep model to retain the original aesthetic information. Other solutions include adaptive spatial pooling~\cite{mai2016composition}.

\citet{kong2016photo} present an \emph{aesthetics dataset} providing an overall aesthetic score and eleven aesthetic attribute scores for over 10,000 images. This paper also proposes a Siamese architecture that estimates the relative aesthetic quality of two input images for better distinguishing images of similar quality. \citet{wang2016brain} predict different style attributes and use them as the input to a final CNN for predicting the overall score distribution. \citet{wu2022interpretable} pushes forward this line of research by introducing interpretability into deep aesthetic models. The author estimates the contribution of each attribute to the overall aesthetic score via learning a hyper-network and estimates the contribution of different image regions to attribute scores by attention mechanism. Similar to \citet{wu2022interpretable}, we also use an attention model, but the model is specially designed to reveal the contribution of cluttered objects to the overall aesthetic appearance and contents. 

Previous approaches work on the image level and evaluate the aesthetic quality of an image as a whole. To the best of our knowledge, the clutter detection model proposed in this paper is the first to study the aesthetic contribution of individual visual elements.

\subsection{Image Inpainting}

In our system, we remove detected clutter from the photo and generate semantically reasonable and visually plausible contents for the missing regions based on background texture. In the field of computer vision, \emph{image inpainting}~\cite{bertalmio2000image} is the technique that synthesizes alternative contents for the corrupted regions~\cite{kim2021learning,li2021faceinpainter,wang2021image,peng2021generating,liu2021pd,liao2021image}. This technique has been extensively studied in many image processing tasks such as image re-targeting, movie restoration, object removal, and old photo restoration~\cite{yu2021wavefill}.

Conventional image inpainting methods exploit neighborhood information for content refilling. Specifically, diffusion methods~\cite{bertalmio2000image,ballester2001filling} propagate neighboring information to the corrupted regions by methods such as interpolation. When the corrupted region is large, there is a lack of global information, and it is hard for diffusion methods to recreate meaningful structures. Patch-based methods~\cite{barnes2009patchmatch,darabi2012image} complete images by searching and reusing similar patches from the background. They work well for images with repetitive contents but struggle with generating meaningful semantics.

Deep learning motivates image inpainting research efforts to shift toward data-driven learning-based approaches~\cite{wu2020leed,wu2020cascade,yu2021diverse,zhan2021bi,zhan2021unbalanced,zhan2019spatial}. Particularly, generative adversarial networks (GANs)~\cite{goodfellow2014generative} have proven their ability to infer contents for corrupted regions with both reasonable structures and realistic appearance~\cite{pathak2016context}. \citet{nazeri2019edgeconnect} supplement GANs with edge detection and salient edge prediction. The idea is that edges provide predictable structural information for corrupted regions. \citet{wang2018image} introduce different receptive fields into image inpainting by utilizing a multi-column network structure. \citet{zeng2019learning} design a hierarchical pyramid-context encoder to use the information at multiple scales for recovering the missing regions. \citet{liu2020rethinking} disentangle structures and textures and recover them separately by deep and shallow features. \citet{liu2018image, yu2019free} propose partial and gated convolution for image inpainting, respectively. Although algorithms based on GANs significantly improve image inpainting performance, they often generate artifacts for large missing regions. \citet{zeng2020high} provide an iterative method to fill the missing regions progressively. At each iteration, they treat inferred pixels with a high confidence value as part of the original images. A drawback of such an iterative method is the relatively low inference speed. In this paper, we borrow the idea of iterative image inpainting and accelerate its speed for fulfilling the needs of real-time interactions.

Related to our work, \citet{fried2015finding} incrementally remove automatically detected distractors by dragging a slider. Our work differs from this system in (1) \citet{fried2015finding} focus on small visual elements while our method is effective for distractors of various sizes; (2) instead of training the distractor detection model using object-level labels as in \cite{fried2015finding}, our clutter detection method uses image-level aesthetic scores generally available in many large-scale datasets.; (3) we additionally develop a capture-time system that visualizes, interactively determines, and removes clutter from photos.

\begin{figure}
    \centering
    \subfigure[In Group 3, the photographer didn't realize some selected subjects (the wine, syrup, and bottles) actually prevented the critic from receiving the intended story. In the second trial, the photographer focused on items on the table and inadvertently included the light switch in the photo, which distracted the mediator. ]{\includegraphics[height=0.278\linewidth]{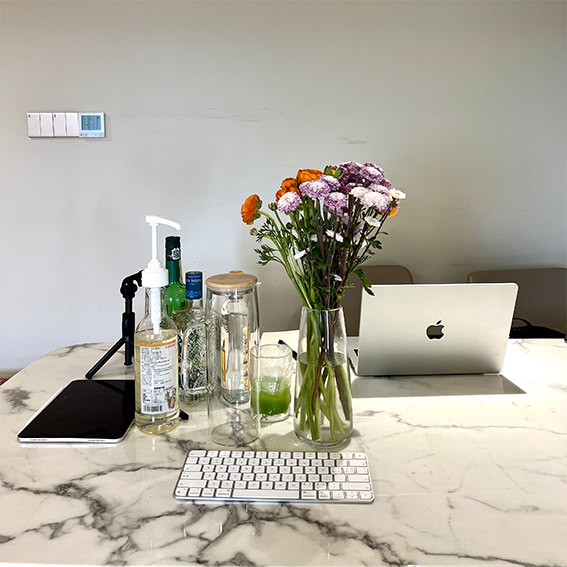}
    \includegraphics[height=0.278\linewidth]{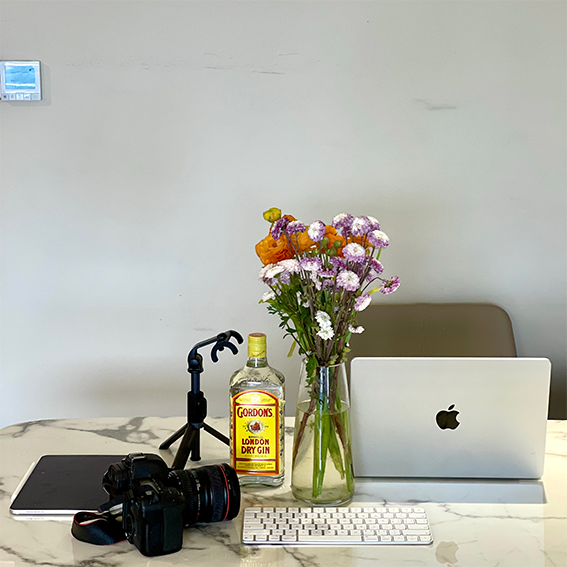}}\hfill
    \subfigure[In Group 6, the photographer became aware of the clutter (vehicles) in the photo, but removing the cluttered elements is difficult because of the heavy traffic.]{\includegraphics[height=0.278\linewidth]{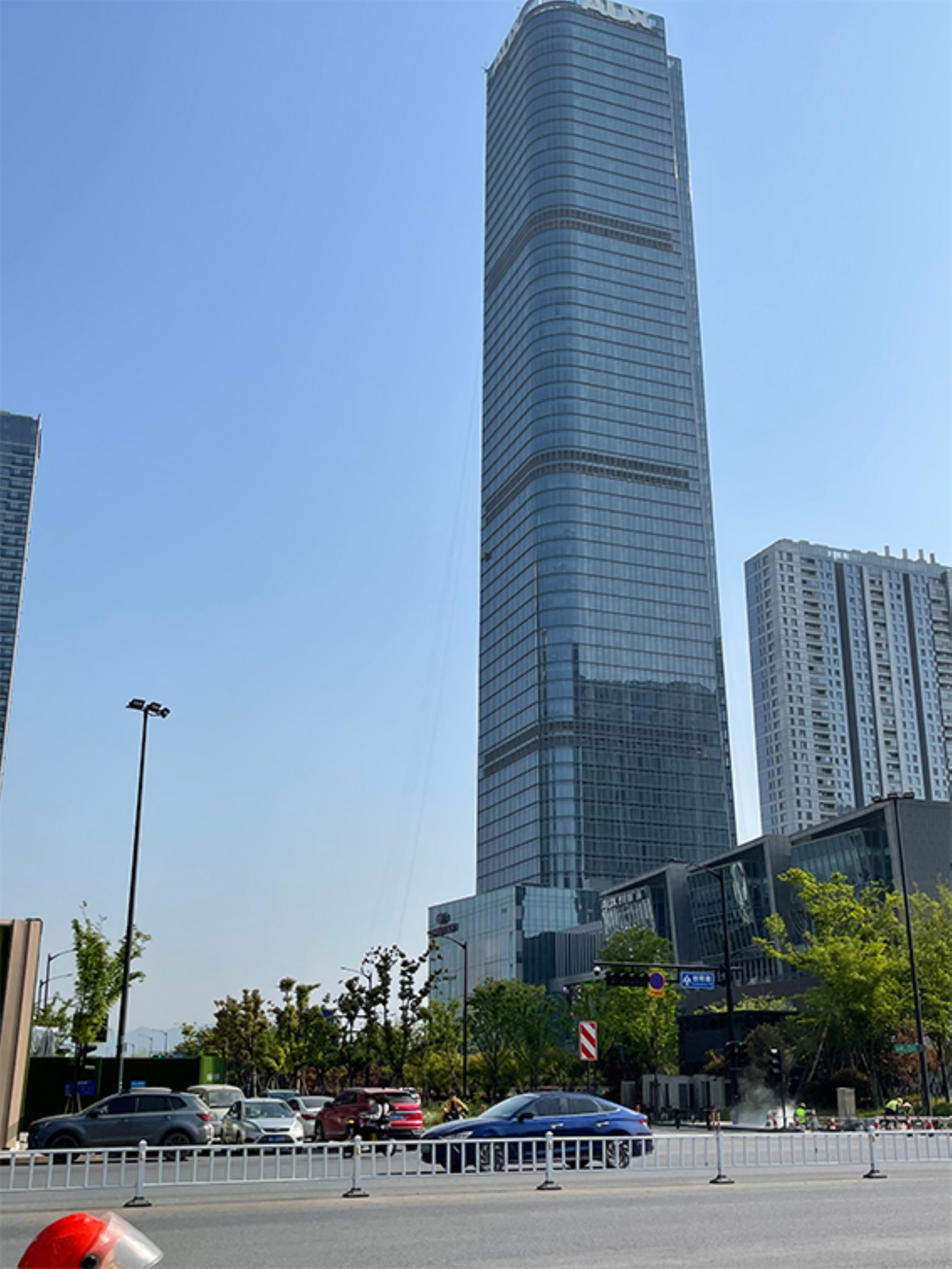}
    \includegraphics[height=0.278\linewidth]{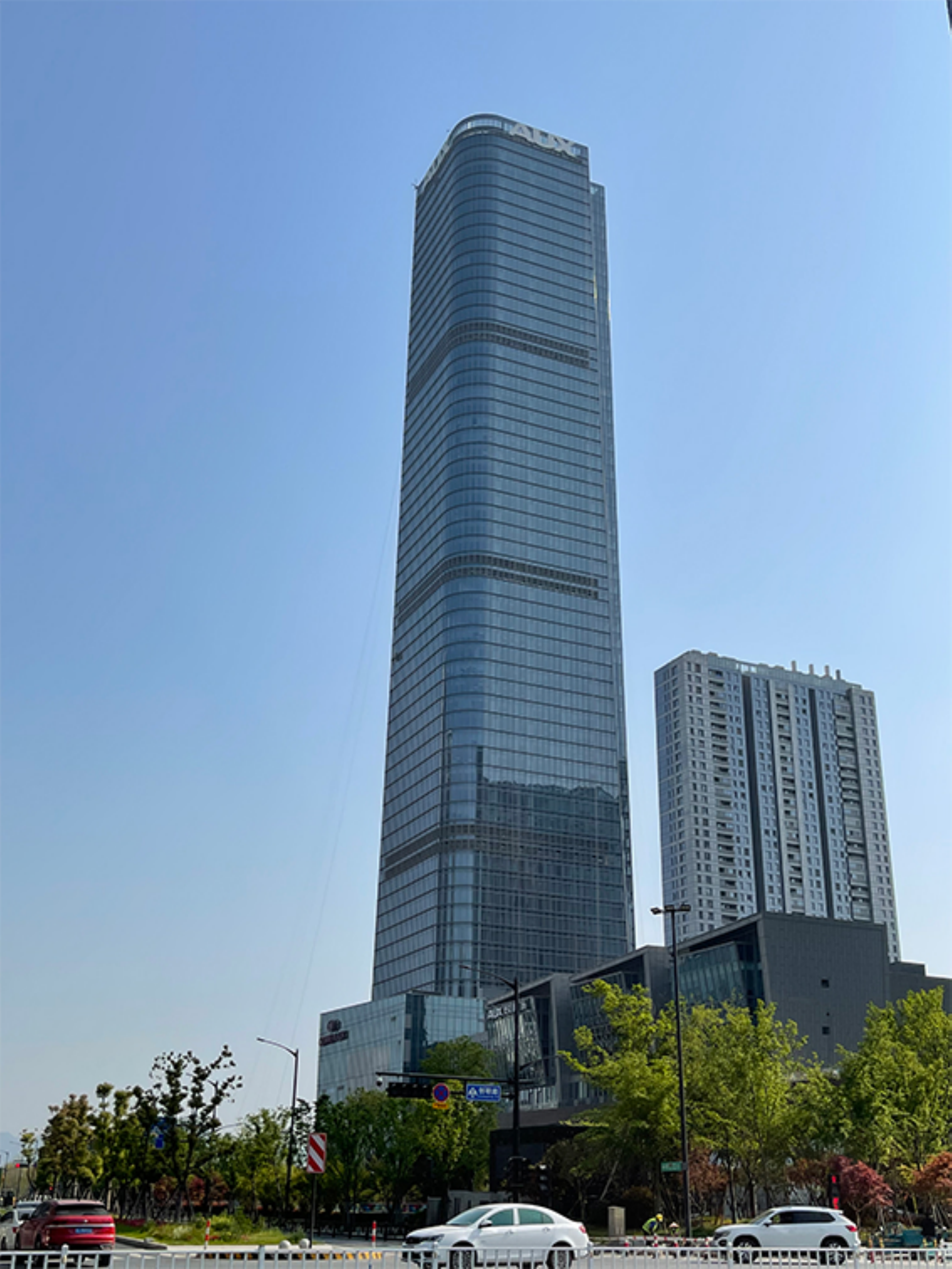}}
    \vspace{-1em}
    \caption{Examples in our pilot study: photos of Group 3 (a) and 6 (b) taken before and after reading the critic's comments.}
    \label{fig:visual_survey}
\end{figure}
\section{Pilot Study}\label{sec:design}

As discussed above, there are photographic guidance systems aiming to remind users of objects in photos. However, the decision of which elements are cluttered is left to users, and the guidance on clutter handling is largely limited. The answers to the following questions are still unclear: (1) What kinds of clutter are common to inexperienced photographers, and how do they influence storytelling? (2) How to help photography amateurs identify visual clutter? (3) How to help users deal with different types of visual clutter to improve photo quality? In this section, we conducted a pilot study to gather information on these problems to help us navigate the design process of our system.


\subsection{Visual Clutter Survey}\label{sec:survey}


\subsubsection{Setup} We recruited 24 participants, with an equal number of females and males. They are aged 23 to 39 years old, with an average age of 28. Since our target users are amateur photographers, we had certain requirements for the background of the participants: (1) The participants did not learn about photography or arts; (2) The participants had empirical photography experience of less than one year. We didn't provide compensation, and the participation was voluntary. The participants are randomly divided into 8 groups, with 3 experimenters in a group. 

The survey goes in the following three steps. (1) Photography. For each group, one of the participants (\emph{the photographer}) is asked to take a photo with the default camera app of the phone. This participant is then asked to give a caption describing the intended story. (2) Feedback. The second experimenter in the group (\emph{the critic}) is tasked to write down its opinion about the photo. Specifically, without knowing the caption of the photographer, the critic tells the story of the photo according to what it sees. The critic is also asked which visual elements support its story. (3) Re-photo. The photographer reads the critic's comments. If the story is significantly different, the photographer can take another photo telling the same intended story. The third experimenter in the group (\emph{the mediator}) sees the two photographs and comments on the stories. 

\subsubsection{Results and Analyses} 

For the first example, the photographer in Group $3$ shot the scene of her home office, including electronic devices and auxiliary equipment for remote working use, as well as daily necessities such as wine and bottles (Fig.~\ref{fig:visual_survey} (a) left). The photographer aimed to document "\emph{the special experience that closely combines life and work during the epidemic.}" However, the critic told a different story: "\emph{My attention was first drawn to the wine, syrup, and bottles in the middle of the scene. Clearly, there is a bottle of mixed drinks there. Then I noticed the iPad and the laptop on the table. The flowers should be for decorative purposes. Based on these observations, I guess the photographer wants to share that she was learning to stir up mixed drinks online}."

Here we see how the clutter prevents the convey of the photographer's intentions. Wine, bottles, and syrup take up a large part of the picture. The photographer thinks that these objects are not so obtrusive because she works here and is familiar with these objects. By contrast, from the eyes of the critic, these objects together draw his attention, and he thinks these objects are the major part of the photo. We see $\mathtt{the\ first\ kind\ of\ clutter}$ in this case: the photographer gets used to some objects and is not aware that they draw much attention and clutter the picture. 

On receiving feedback from the critic, the photographer became more intentional about staging the elements critical to her stories. She removed many bottles and zoomed in to emphasize the home scene (Fig.~\ref{fig:visual_survey} (a) right). We found that, to get rid of clutter, many (4) photographers chose to remove objects according to the critic's comments, while others chose to zoom in (2), adjust the lighting of the scene (2), or change orientation from portrait to landscape (1).

The mediator successfully got the intention of the photographer and guessed that she was framing a home working scene. However, the mediator was distracted by some other visual elements when he saw the photo for the first time: "\emph{My eyes stopped on the light switch for a long time. I know it was not the critical part, but I just cannot help myself. I started to see other parts of the photo only after I could tell the symbol on the switch}." The light switch is a representative example of $\mathtt{the\ second\ kind\ of\ clutter}$ -- the photographer is focusing on other aspects of the photo and ignores some visual elements that are irrelevant to the story of a photo but are obtrusive.

For the second example, the photographer of Group $6$ (Fig.~\ref{fig:visual_survey} (b) left) framed a street scene with a magnificent building to express "\emph{the joy of getting a new job in a beautiful building}." However, the critic thought: "\emph{I guess the photographer is documenting a street scene of the building, but the vehicles and street lights clutter the photo, some of which are not visually pleasant. Elements like these take the attention away from the subject of the photo.}" Based on the comments of the critic, the photographer sought to get rid of the clutter in the second attempt. However, he found it difficult because the traffic was heavy and there was a construction site that was very close to the building. Therefore, as shown in Fig.~\ref{fig:visual_survey} (b) right, the second attempt was still cluttered with visual distractions. The mediator said: "\emph{From the comparison of these two photos, the photographer tried to reduce clutter in the scene. It is not easy. However, this situation is quite common -- just imagine you are taking photos at a resort with many tourists.}" The mediator's comments reveal $\mathtt{a\ third\ kind\ of\ clutter}$: the photographer is aware of the clutter, but it is hard to remove it from the scene.

\subsection{Survey Insights and Design Goals}\label{sec:design_goal}

The survey presented in the previous sub-section gives the following key insights (KI).

\emph{KI1: Photographer amateurs commonly encounter three kinds of clutter in capture time}. (1) The photographer is familiar with the photo scene, being unaware of visual elements that are obtrusive to the audience. (2) The photographer ignores some visual distractions when focusing on other aspects of the photo. (3) The photographer knows the clutter, but it is difficult to remove it. 

\emph{KI2: Users have different needs for handling these types of clutter}. Inexperienced photographers struggle to identify the first kind of clutter because the gap between this kind of clutter and meaningful content is narrow. Users need hints or guidance on distinguishing this type of clutter. Dealing with the second kind of clutter is relatively easy. We can highlight the neglected clutter to remind the photographer. For the third kind of clutter, users need an interactive tool that can remove the clutter and fill the missing regions with reasonable background content.

\emph{KI3: Users value capture-time guidance and feedback regarding clutter detection and handling.} Seven (out of eight) users believed they could take a better photo if the suggestion about which elements are distracting and the feedback about how their photos look after removing clutter were available during shooting. For example, $P3$ said: "\emph{I wish the critic were by my side when taking pictures. He would let me know what my picture looks like to other people.}" $P6$ said: "\emph{When I get home, I will use some post-processing tools to remove the vehicles. I am worried about that. If my picture looks bad after removing these objects, I would have to return to the spot for another shoot.}"

\emph{Design Goals.} On the basis of prior work and our findings in the pilot study, we derived three major goals for the design of our capture-time clutter-handling system. Goal 1 interactively reminds the photographer of distracting elements in the photo. This goal deals with the first two types of clutter. Goal 2 gives the user suggestions about how to remove clutter. Goal 3 provides a capture-time preview for clutter removal so that the user can fix possible problems on the spot. This third goal targets at the third type of clutter.

These design goals inspire the interaction design of our system described in Sec.~\ref{sec:system_overview} and the technical novelties that support these interactions introduced in Sec.~\ref{sec:method-aes} and~\ref{sec:method-removal}.

\vspace{-1em}
\section{DeclutterCam}\label{sec:method}
\subsection{System Overview}\label{sec:system_overview}

Our system is a camera app with standard photographing functions and several novel interactions related to clutter detection and removal inspired by the investigations in our pilot study. On opening our system, the user sees a viewfinder with \emph{masks highlighting visual elements} (Fig.~\ref{fig:interaction} (c)) in the scene. On this default user interface, detected visual clutter is emphasized with a bright color while normal elements appear with a transparent mask. The underlying clutter detection algorithm (described in Sec.~\ref{sec:method-aes}) gives an estimation $q$ for each element \emph{quantifying\ the\ contribution\ of\ this\ element to the overall content and aesthetic quality}. The contribution $q$ can be positive or negative, with a negative value indicating that the element lowers the overall quality and may be clutter. We classify elements with negative $q$ values as clutter. \emph{By\ clicking\ an\ element,\ the\ user\ sees\ the\ contribution\ value\ $q$\ and\ removal\ suggestions\ for\ clutter} (Fig.~\ref{fig:interaction} (b)). The masks and contribution values can help deal with the first two kinds of clutter discussed in Sec.~\ref{sec:survey} -- the user now gets aware of the ignored cluttered elements.


There are two kinds of suggestions for removing clutter. The first type includes conventional photographic tricks: zooming in, changing camera position, and changing orientation from portrait to landscape. The second type of suggestion is to use our image inpainting tool. The first type of suggestion is given only when following it will not significantly change the original photo's composition. In practice, we give this type of suggestion when the element is near the photo boundary and its area is smaller than a threshold $s_\tau$.

If the user wants to see the original scene, \emph{the masks become invisible when the user clicks the "eye" button} (Fig.~\ref{fig:interaction} (a)). Moreover, if the classification results of our detection algorithm conflict with the user's judgment, the user can \emph{change the category of an element by\ double-clicking\ it.} (Fig.~\ref{fig:interaction} (d)). This kind of conflict reflects a concept in the field of computational aesthetics. Data-driven approaches to aesthetics like ours try to reconcile the conflict between subjectivism and objectivism~\cite{zuckert2007kant} by assuming that judgments provided by a pool of human observers approximate the true aesthetic value. We train our clutter detection model via regression to such a pool of judgments. In this way, we expect it to identify visual elements that are commonly perceived as clutter by most observers but ignored by photography beginners.

The user can remove the cluttered elements according to the suggestions of our system. For those cluttered elements that are not easy to be removed, \emph{the user can click the "clean" button to trigger the image inpainting algorithm} (Sec.~\ref{sec:method-removal}) \emph{and preview the scene without these elements} (Fig.~\ref{fig:interaction} (e)). On clicking this button, the back end will remove the elements that are currently classified as clutter. With preview images as timely feedback, the user can find possible problems of the cleaned photo, e.g., unbalanced composition or inharmonious color distribution. Adjusting their photo and avoiding these problems in capture-time will help the user navigate the photographic design process and take photos of higher quality right on the spot.



The realization of these interactions motivates us to develop computational modules for the following two functions: (1) classifying normal and cluttered elements by calculating their contribution to the overall content and aesthetic quality; (2) removing the cluttered elements and reconstructing the missing regions with realistic background context. In the next sections, we describe the technical details of the realization of these functions.

\subsection{Clutter Identification}\label{sec:method-aes}

\begin{figure}
    \centering
    \includegraphics[width=\linewidth]{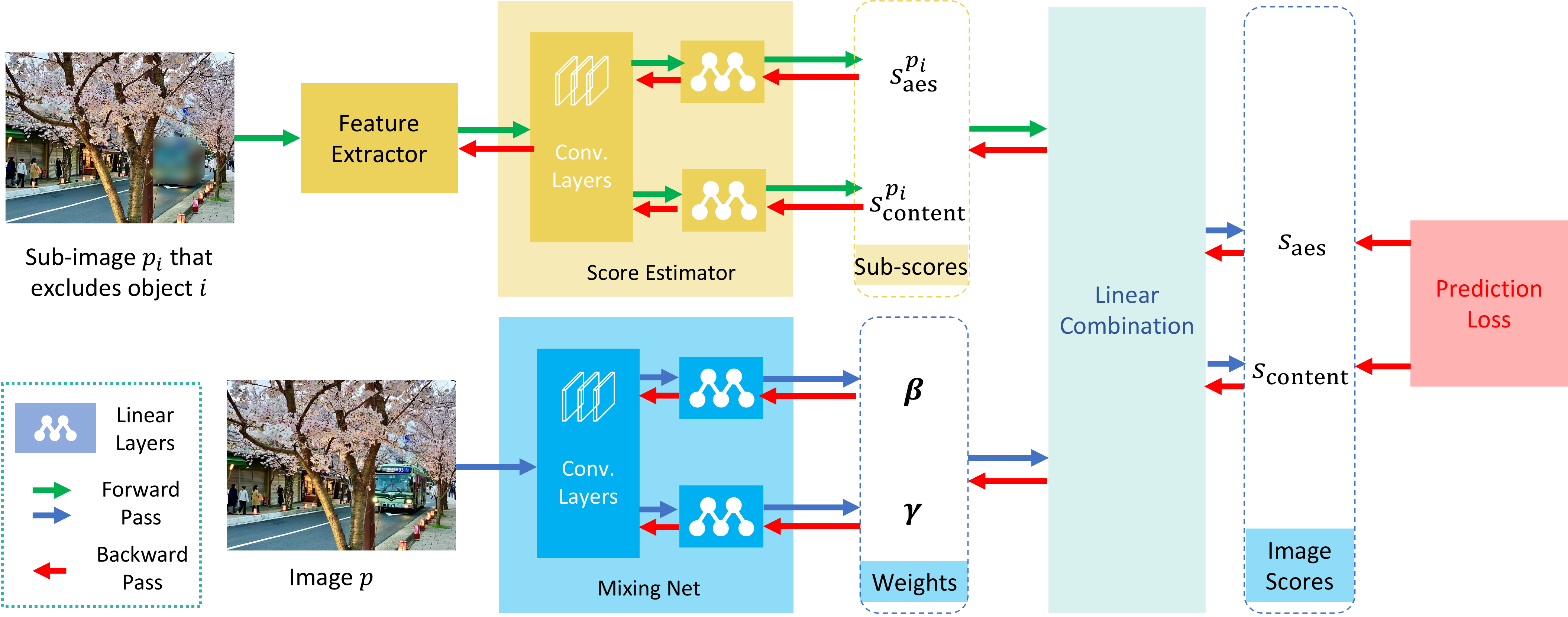}
    \caption{Network architecture of our framework that estimates the contribution of elements to the overall aesthetic quality and content. }
    \label{fig:aes-framework}
\end{figure}
In this section, we discuss how to distinguish cluttered elements. We first provide a quick overview of our method. We run an object detection algorithm to find out all visual elements and treat them as clutter candidates and then use a novel evaluation model to estimate the contribution of each element to the photo's overall aesthetic and content quality. Elements with negative contribution values are classified as clutter.

The first step, object detection, is extensively studied in the field of computer vision. Mask R-CNN~\cite{he2017mask} is a classic and accurate deep model that predicts the object labels and provides pixel-level masks showing the detailed position of objects (instance segmentation). Formally, given an input image of the size $h\times w$, we use the instance segmentation head of a trained Mask R-CNN model that outputs $k$ masks $\{m_1, m_2,\dots,m_k\}$ for $k$ detected elements. Each mask $m_i\in\mathbb{R}^{h\times w}$ has the same size as the input image. Masks are binary, with $1$ meaning that the corresponding pixel belongs to the element and $0$ meaning the opposite.

Mask R-CNNs are typically trained on the COCO dataset~\cite{lin2014microsoft}. This dataset contains $80$ categories of objects common in daily life. A question is that clutter does not always manifest itself as objects and can appear as stains, dirt, and other non-object structures. To make our system generally applicable to more types of clutter, we add two new categories to the COCO dataset to train the Mask R-CNN. The first category is for line-shaped clutter, and we label power lines, plumbing, and wires. The second category is for clutter in irregular shapes, including oil/water-based stains, varnish/lacquer stains, and dirt. We followed the labeling process of the COCO dataset~\cite{lin2014microsoft}, and each of the new categories has a similar number (5000) of labeled instances with original categories.


With the detected elements, we now introduce a novel deep neural network model that learns the contribution of each element to the overall content and aesthetic quality.

Modern image aesthetic datasets, like AADB~\cite{kong2016photo}, provide scores, $s_{\text{aes}}$, and $s_{\text{content}}$, for each image reflecting whether the image is aesthetically appealing and contains interesting content, respectively. Our idea here is to learn these overall scores as a decomposition of scores for sub-images that exclude given elements. Based on this decomposition, we can calculate the contribution of each element.

The proposed learning framework (Fig.~\ref{fig:aes-framework}) consists of three modules. The first part is a pre-trained ResNet~\cite{he2016deep} as a feature extractor. Input to this feature extractor is a sub-image that blurs a given element. Specifically, for each element mask $m_i$ given by the object detection algorithm, we generate a sub-image 
$p_i$ by performing Gaussian blurring to the regions where the value of $m_i$ is 1. The blur is processed by calculating the convolution of image patches with a $13\times 13$ Gaussian kernel with a variance of 1. The output of the feature extractor is feature maps $z_{p_i}$ which are then fed into a score estimator. The score estimator outputs two scores for each sub-image. The first score $s^{p_i}_{\text{aes}}$ is an estimation of the aesthetic quality of the input, while the second score $s^{p_i}_{\text{content}}$ estimates the quality of content. The score estimator is a convolutional neural network with learnable parameters $\theta_{\text{score}}$. This network has two convolutional layers, a flattened layer, and two output heads. Each output head is a two-layer fully-connected network. We denote the score estimator by $f_{\text{score}}$, and $(s^{p_i}_{\text{aes}}, s^{p_i}_{\text{content}}) = f_{\text{score}}(p_i; \theta_{\text{score}})$.

Although we now get the scores related to individual elements, how to train $f_{\text{score}}$ is a problem. Ground-truth scores are provided for images, and we do not have element-level labels, meaning that we lack supervision signals for $f_{\text{score}}$. To solve this problem, we propose to learn the overall score as a combination of individual scores:
\begin{equation}
    s_{\text{aes}} = \sum_{i=1}^k \beta_i * s^{p_i}_{\text{aes}},\ \ \     s_{\text{content}} = \sum_{i=1}^k \gamma_i * s^{p_i}_{\text{content}},
\end{equation}
where $s_{\text{aes}}$ and $s_{\text{content}}$ are the overall aesthetic score and content prediction, respectively. Then the model can be updated by minimizing two prediction losses. The first one is the mean square error between the predicted aesthetic quality $s_{\text{aes}}$ and the ground-truth $y_{\text{aes}}$:
\begin{equation}
    \mathcal{L}_{\text{aes}}(\theta_{\text{score}}, \theta_{\text{mix}}) = \mathbb{E}_{p\sim\mathcal{T}}\left[\left(y_{\text{aes}}(p) - s_{\text{aes}}(p;\theta_{\text{score}}, \theta_{\text{mix}})\right)^2 \right],\label{equ:loss_aes}
\end{equation}
where the expectation operator means that images $p$ are sampled from the training set $\mathcal{T}$. The second loss is the content score prediction error given by:
\begin{align}
    \mathcal{L}_{\text{content}}(\theta_{\text{score}}, \theta_{\text{mix}}) = \mathbb{E}_{p\sim\mathcal{T}}\left[\left( y_{\text{content}}(p) - s_{\text{content}}(p;\theta_{\text{score}}, \theta_{\text{mix}})\right)^2 \right],\label{equ:loss_content}
\end{align}
where $y_{\text{content}}$ is the ground-truth score for the content available in the dataset. By introducing a scaling factor $\lambda_{\text{aes}}$, the total loss for training our model is: $\mathcal{L}(\theta_{\text{score}}, \theta_{\text{mix}}) = \lambda_{\text{aes}} \mathcal{L}_{\text{aes}}(\theta_{\text{score}}, \theta_{\text{mix}}) + \mathcal{L}_{\text{content}}(\theta_{\text{score}}, \theta_{\text{mix}}). \nonumber \label{equ:total_loss}$

One question of this model is how to get the weights, $\beta$ , , and $\gamma$. Intuitively, these weights reflect the interrelationship between elements and should be determined by the structure of the original image. Therefore, we adopt a network conditioned on the input image to learn the weights. Specifically, a network $f_{\text{mix}}(\ \cdot\ ; \theta_{mix})$ with two output heads, each with a SoftMax activation after the last layer, processes the input image and outputs the vectors $\bm \beta=\langle \beta_1, \beta_2, \dots, \beta_k\rangle$ and $\bm \gamma=\langle \gamma, \gamma_2, \dots, \gamma_k\rangle$. It is worth noting that gradients can flow through $\bm \beta$ and $\bm\gamma$ into $f_{\text{mix}}$, and can also flow into the score estimator through $s_{\text{aes}}$ and $s_{\text{content}}$. Therefore, our learning framework is end-to-end differentiable and can be trained as a whole by minimizing the prediction losses.

With a trained score estimator, we can now distinguish cluttered elements from normal ones. For the $i$th element in the scene, we obtain its aesthetic score $s^{p_i}_{\text{aes}}$, content score $s^{p_i}_{\text{content}}$, and the corresponding weights, $\beta_i$ , and $\gamma_i$, by running our model. The semantic meaning of these scores is the quality of the image without the $i$th element. With the estimated overall scores for the original image, we can calculate the contribution of the $i$th element to the whole image:
\begin{align}
    q_i = \beta_i(s_{\text{aes}} - s^{p_i}_{\text{aes}}) + \gamma_i(s_{\text{content}} - s^{p_i}_{\text{content}}).
\end{align}
A negative value of $q_i$ indicates that the $i$th element lowers the overall quality of the photo. Therefore, we classify these elements as clutter. If the classification conflicts with the impression of the user, it can change the category of an element by double-clicking it on the scene.

\subsection{Clutter Removal}\label{sec:method-removal}
The second function that we need to support is to remove the cluttered elements and fill the missing regions with realistic background context. The cluttered elements may be detected by our algorithms or selected by the user, and we can determine their positions by their masks $\langle m_{c_1}, \dots, m_{c_n} \rangle$, where $c_1, \dots, c_n$ are the indices of the cluttered elements and are in the set $\{1, \dots, k\}$. We assume that there are $n (\le k)$ cluttered elements in the photo.

As discussed in the related work section, image inpainting is the technique that fills missing areas with background content. Generative adversarial networks (GANs) are frequently used to finish this task but are less efficient for high-resolution images. For comparison, state-of-the-art image inpainting algorithms (e.g.,~\citet{guo2021image,zeng2021generative}) works on $256\times 256$ images while photos taken by iPhone 13 camera app is of the size $4032\times 3024$.

The idea of solving the problem is to iteratively fill the missing area. An additional, lightweight branch predicting which generated pixel is more like an artifact is added. In each iteration, we fill the part of the missing region with high-quality pixels and treat the other parts as the new missing region and run the next iteration. We note that~\citet{zeng2020high} use a similar iterative approach and find that it can improve the image quality in filled regions. Our method can be regarded as a lightweight and faster version of~\cite{zeng2020high}.


The input to the image inpainting model is the image with a clutter mask $m_c$. We denote this corrupted image as $p_c=p \circ (1-m_c)$, where $\circ$ is the element-wise multiplication. The first branch of the generator $G$, parameterized by $\theta_g$, outputs the inpainted image: $y=G(p_c)$, while the second branch, parameterized by $\theta_{b}$, gives a probability map $b$ indicating how likely each pixel is an artifact. The model is trained by a reconstruction loss and an adversarial loss:
\begin{align}
    \mathcal{L}_G(\theta_g) = \mathbb{E}_{p\sim\mathcal{T}}\left[\|y-p\|_1 + (1-D(p\circ(1-m_c) + y \circ m_c))\right].
\end{align}
Here, $\mathcal{T}$ is the dataset for training, $\|y-p\|_1$ is the reconstruction loss, and $D$ is the discriminator parameterized by $\theta_d$. ($D$ has two outputs, with $1$ meaning a real image and $0$ meaning a fake image.) The generator is optimized to enforce the output of $D$ to be 1, meaning that the discriminator thinks $y$ is very likely to be a real image. In the meantime, the discriminator is trained to minimize
\begin{align}
    \mathcal{L}_D(\theta_d) = \mathbb{E}_{p\sim\mathcal{T}}\left[1 - D(p)\right] + \mathbb{E}_{p\sim\mathcal{T}, y=G(p)}\left[1 + D(p\circ (1-m_c) + y \circ m_c)\right],
\end{align}
to be able to distinguish the generated images from the real images in the training dataset. The second branch of the generator is updated by minimizing:
\begin{align}
    \mathcal{L}_b(\theta_b) = \mathbb{E}_{p\sim\mathcal{T}}\left[ m_c \circ((1-b)\circ |y-p|)\right] .
\end{align}
By minimizing this loss, $b$ value is high only when the generated pixel in the missing regions is far away from the ground-truth pixel. Then we select regions with low $b$ values as the missing regions for the next iteration. To guarantee that our system responds on time, we set the maximum iteration time to 5 for capture-time guidance. For the detailed training scheme of our models, please refer to Appendix~\ref{sec:training_scheme}.

\section{User Evaluation}\label{sec:exp-com}

In this section, we carry out user studies to better understand participants' performance and behavior using \name. Specifically, we are interested in (1) whether users think our clutter detection and removal algorithms are accurate; (2) whether our system can help users better handle clutter and get photographic works of higher quality; and (3) whether our system provides user-friendly interactions and interfaces. To answer these questions, our user study consists of two major parts. The first part is for testing functional sub-modules of clutter detection and removal, and the second part is for the system as a whole. 

In these studies, we recruited 32 participants (16 male, 16 female) aged 19 to 55 years old ($\mu$=34). Since our system is mainly designed for photography beginners, we placed restrictions on the background of participants: (1) The participants did not learn about photography or arts; (2) The participants had empirical photography experience of less than one year. Each participant was compensated \$5 for taking part in the study. Before the studies, we helped the participants install and get familiar with our app. The participants were encouraged to share their thoughts and usage experience with us anytime during the experiments. 


\subsection{Study Procedure}\label{sec:exp-setup}

\subsubsection{Modular Tests} To know the reaction of users to our clutter detection and removal algorithms, we design two user studies. 

The first user study evaluates our clutter detection sub-module, including the classification of clutter and normal elements and the contribution of each element to the overall photo quality. Experimenters were invited to select a scene and take a picture. After that, each participant was randomly assigned three photos of other participants, and the task is to (1) label visual elements in these photos as clutter / not-clutter without seeing the predictions of the algorithm; (2) give each element a score quantifying the estimated contribution of the element to the overall image quality. 

For quantitative evaluation, we (1) compared the user classification results to the predictions of our algorithm by counting the number of elements where the predictions conflict; (2) counted the number of user-labeled clutter elements that were missed by our algorithm; (3) compared the contribution estimation given by participants and the algorithm. For qualitative evaluation, we asked participants to answer a Likert question on a 7-point scale about their impression of the accuracy of clutter detection. We also conducted an open-ended interview with participants talking about their thoughts about the algorithm performance and the design of this interaction. 

In the second user study, users were tasked to click the "clean" button and evaluate the image inpainting algorithm. They answered three Likert questions on a 7-point scale: (1) whether the generated content is visually plausible? (2) whether the generated content is semantically reasonable? (3) whether the inpainted image is of high overall quality? (4) how frequently do you use the removal function during shooting? We compared our clutter removal algorithm against a baseline image inpainting algorithm~\cite{guo2021image} that has achieved state-of-the-art performance on various datasets like CelebA~\cite{liu2015deep}, Paris StreetView~\cite{doersch2012makes} , and Places2~\cite{zhou2017places}. We also conducted an open-ended interview in this study, exchanging opinions about the performance of the algorithm with the experimenters.

\begin{figure}
    \centering
    \includegraphics[width=\linewidth]{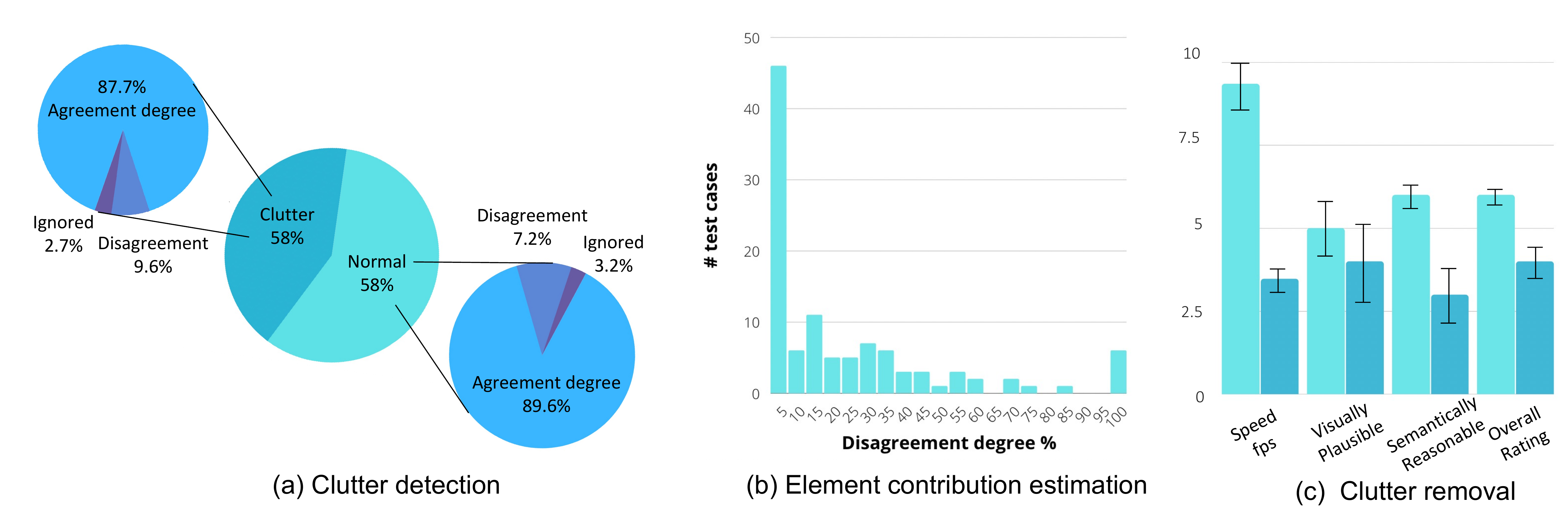}
    \vspace{-2em}
    \caption{Quantitative evaluation of modular tests. (a) Compared to human-labeled results, the agreement rate and missing rate of our clutter detection module. The overall agreement rate is $91.88\%$, and the overall missing rate is $2.99\%$. (b) The distribution of disagreement degrees between human-labeled results and our contribution evaluation module's predictions. For example, in 11 test cases, the disagreement rates are in the range $(10\%, 15\%]$. The overall disagreement rate is $9.91\%$. (c) Comparison between our clutter removal module and the baseline image inpainting baseline~\cite{guo2021image} regrading the time-efficiency and quality (visual plausibility, semantic reasonableness, and overall quality).}
    \label{fig:quantitative_1}
\end{figure}
\subsubsection{System Tests} 

In order to evaluate our system, we compared against three baselines: (1) the default camera app without any clutter detection or removal function; (2) a system where the clutter detection module is ablated -- experimenters check their photos, circle clutter elements, and remove them using our image inpainting algorithm; (3) a system where the image inpainting module is ablated -- participants adjust their photos according to our clutter detection results. The third baseline is similar to previous work on photography guidance~\cite{jane2021dynamic} which reminds the users of objects in the scene, but with an additional module distinguishing clutter from normal elements. Comparing our system against the first baseline can verify whether the proposed interactions are effective for improving picture quality in a user-friendly manner. The comparison against the second baseline will demonstrate how clutter elements ignored by photography beginners influence the story-telling aspect of the photo. Comparing with the third baseline can verify the role of real-time clutter removal in the exploration of photography ideas. Experience of using these baseline systems and our system can also test how users react to our interaction and visualization design.


The 36 participants were tasked to take pictures using the baseline systems and our system. We invited three photography experts (two photography graduate students aged 24 and 25, respectively, and a college photography teacher aged 49), as well as the participants, to evaluate these photos. Each participant was randomly assigned photos of three other participants, and the experts were requested to review the photos of all experimenters. We presented four pictures (3 for baselines, 1 for our system) of a participant to reviewers in a random order, and requested them to rate each photo, on a 7-point scale, from four perspectives: (1) \emph{Story}: whether the photo successfully delivered the story or emotion that the participant intended to express. Participants were requested to write down their intentions before the review starts, and reviewers compared them to their impressions after seeing the photos. (2) \emph{Content}: whether the photo contained interesting content. (3) \emph{Clutter}: whether there were distracting visual elements in the photo. (4) \emph{Overall}: the quality of the photo as a whole.

Moreover, experimenters were asked to answer four Likert questions, also on a 7-point scale, about whether they like the interactions (including the actions for clutter removal and requesting suggestions), the visualization, and the overall guidance provided by our system as compared to baseline systems. Both the experimenters and the experts were invited to take an interview where they could freely talk about their experience of using our app and if any, their suggestions for improving the system.

\begin{figure}
    \centering
    \includegraphics[width=\linewidth]{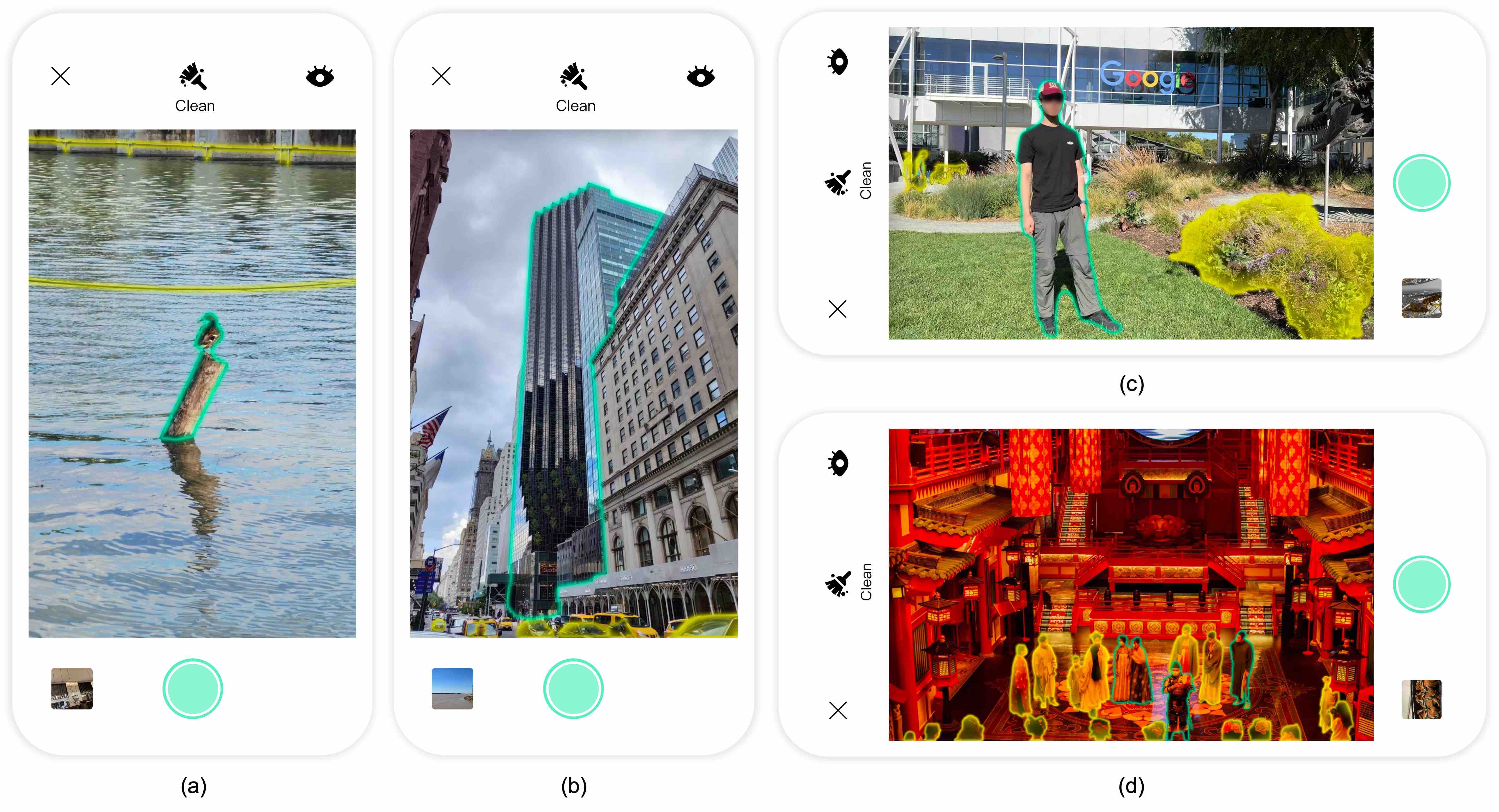}
    \caption{Examples of clutter detection results. (a) $P19$ found that our system successfully recognized the salient object and detected clutter elements in various shapes. (b) $P3$ found that the detection module satisfyingly reminded him of ignored clutter. (c) In $P31$'s photo, our clutter detection module successfully distinguished human objects with different contributions to the image. (d) $14$ wants to include the people. However, our detection algorithm and two (out of three) peer reviewers think too many people are distracting.}
    \label{fig:detection}
\end{figure}
\subsection{Modular Test Results}

\subsubsection{Clutter Detection} We observe clear results that favor our clutter detection results (Fig.~\ref{fig:quantitative_1} (a)). For all the $36$ photos and $108$ test cases (recall that each participant reviewed three photos of other participants), experimenters labeled 1071 visual elements, 408 of which were classified by participants as clutter. Among the 663 participant-labeled normal elements, our algorithm thought 48 ($7.24\%$) of them should be clutter and ignored 21 ($3.17\%$) of them. For the cluttered elements labeled by the participants, 11 ($2.70\%$) of them are ignored and the category for 39 ($9.56\%$) of them was flipped by the algorithm. Overall, our algorithm found $97.01\%$ elements and achieved an agreement degree of $91.88\%$ with the judgment of human experimenters. From an individual perspective, 7 participants did not alter any category, and $P1$ made the most (11) modifications. 

$P19$ finds our detection algorithm accurate and applicable to elements of various shapes (Fig.~\ref{fig:detection} (a)): "\emph{It is interesting to see a bird standing on a wooden stick in the river. I am happy that the system thinks so and gives main objects a high contribution estimation. Meanwhile, the system detects the electric wire and the plumbing, which are of very different shapes, as clutter. The results are in line with my feelings.}" The photographer of this picture claimed that: "\emph{I was unaware of the plumbing during shooting, may be because I had been in the environment for quite a while.}" This example demonstrates that our system can detect the first kind of clutter discussed in Sec.~\ref{sec:design}.

$P3$ reported that the masks drew his attention to clutter which was ignored beforehand (Fig.~\ref{fig:detection} (b)): "\emph{The striking masks caught my eyes. I didn't realize the cars were in the picture.}" $P3$ then used the trick of zooming in to focus on the major object and got rid of the vehicles. The experience of $P3$ indicates that our system meets the expectation for dealing with the second kind of clutter discussed in Sec.~\ref{sec:design}.

$P31$ was also satisfied with the algorithm's prediction results because it successfully distinguished the contribution of different objects of the same type (Fig.~\ref{fig:detection} (c)): "\emph{I wanted take a portrait in front of the building but there were always some other people in the scene. I am glad to see that the algorithm assigns a positive value to the main person and negative values to other people.}"

$P14$ (Fig.~\ref{fig:detection} (d)) found some conflicts between the algorithm's predictions and his judgment. "\emph{I want to include the people. However, the system thinks they are cluttered. I somewhat agree that there are too many people around the stage, but I think my photo becomes less vivid without them.}" Two (out of three) peer reviewers agree with our predictions and think fewer people will make the photo look better. There is a conflict between subjectivism and objectivism in this case, and out system identified visual elements that are commonly perceived as clutter by most observers.


\begin{figure}
    \centering
    \subfigure[]{\includegraphics[height=0.28\linewidth]{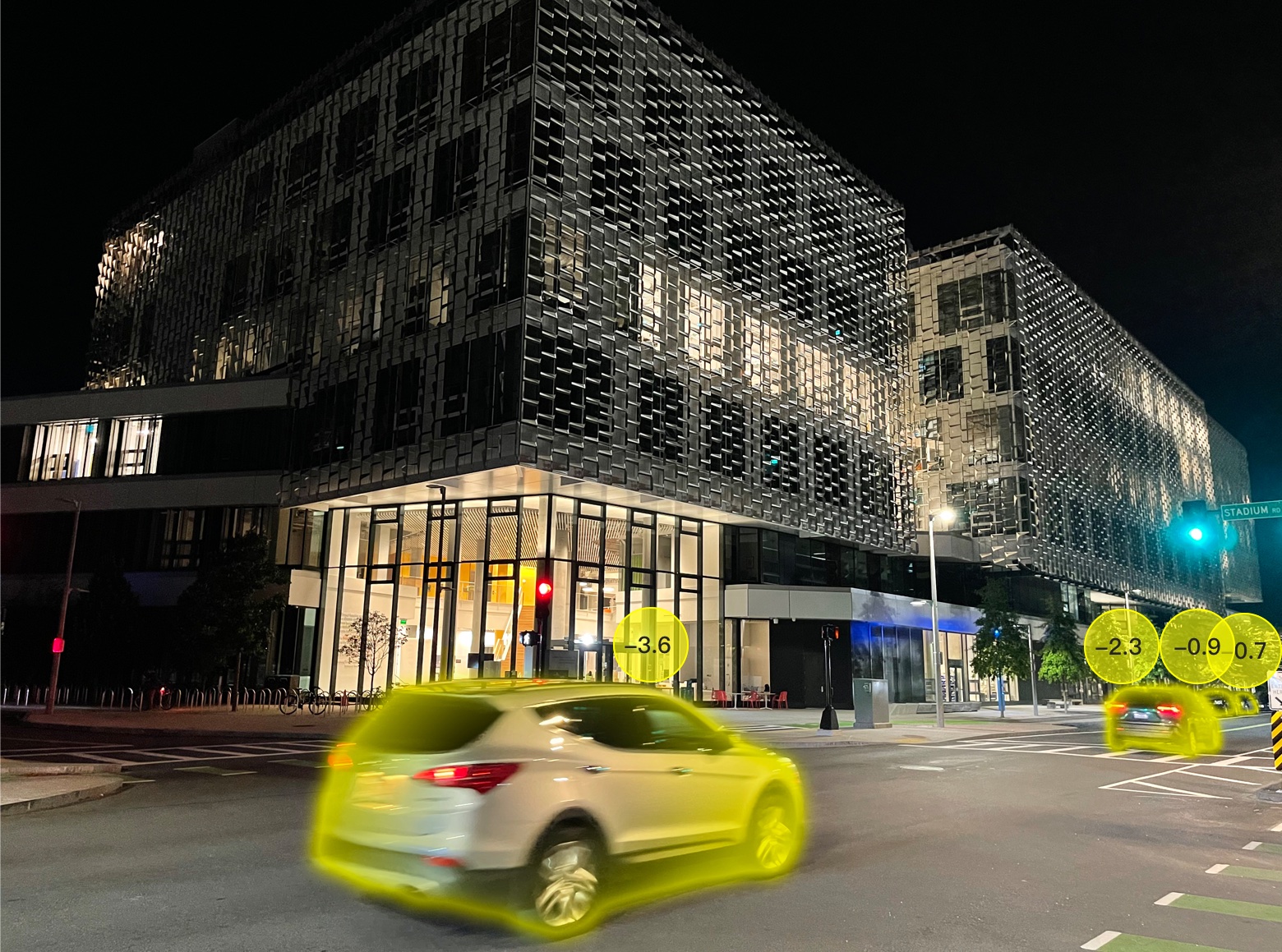}}
    \subfigure[]{\includegraphics[height=0.28\linewidth]{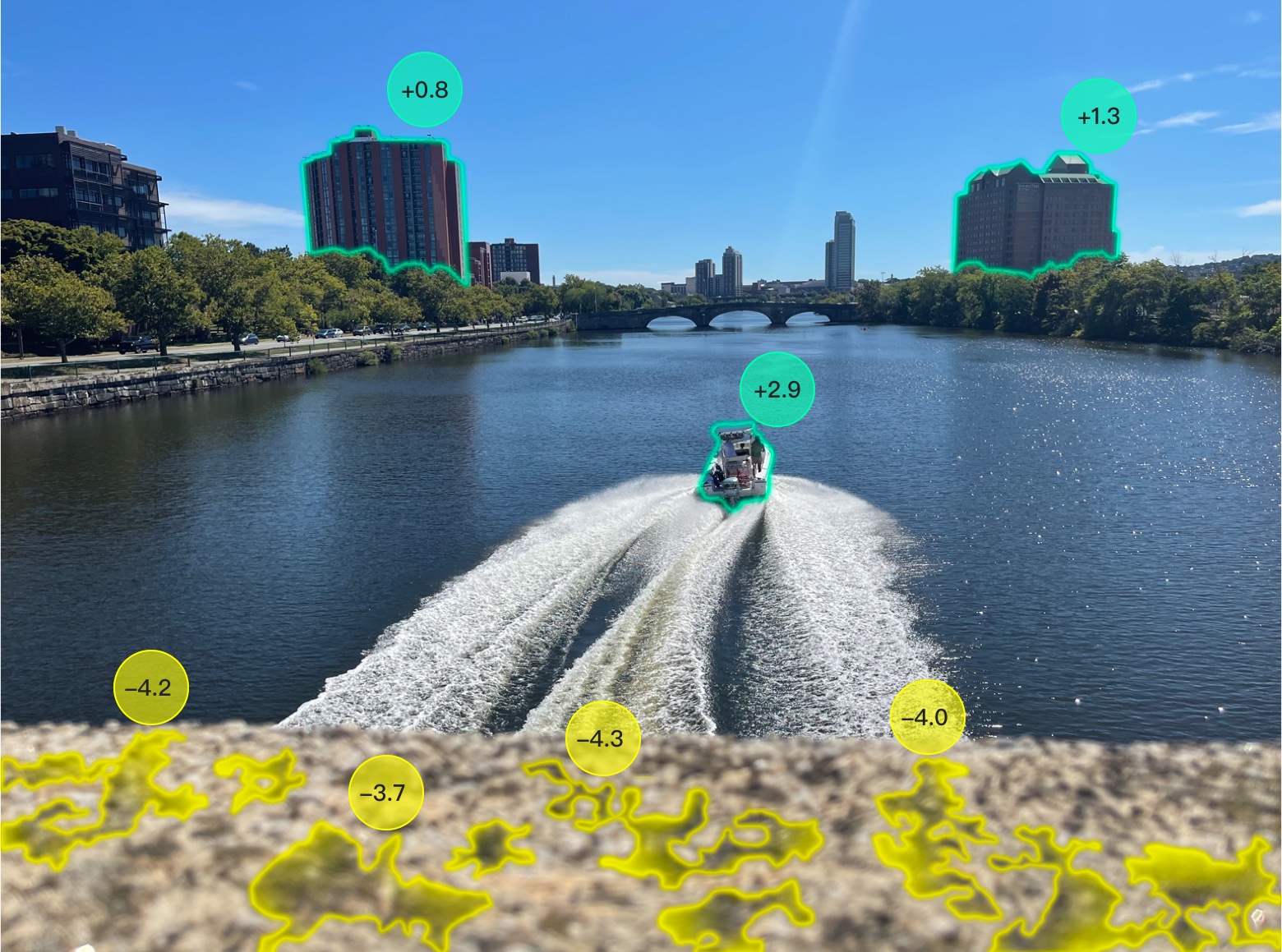}}
    \subfigure[]{\includegraphics[height=0.28\linewidth]{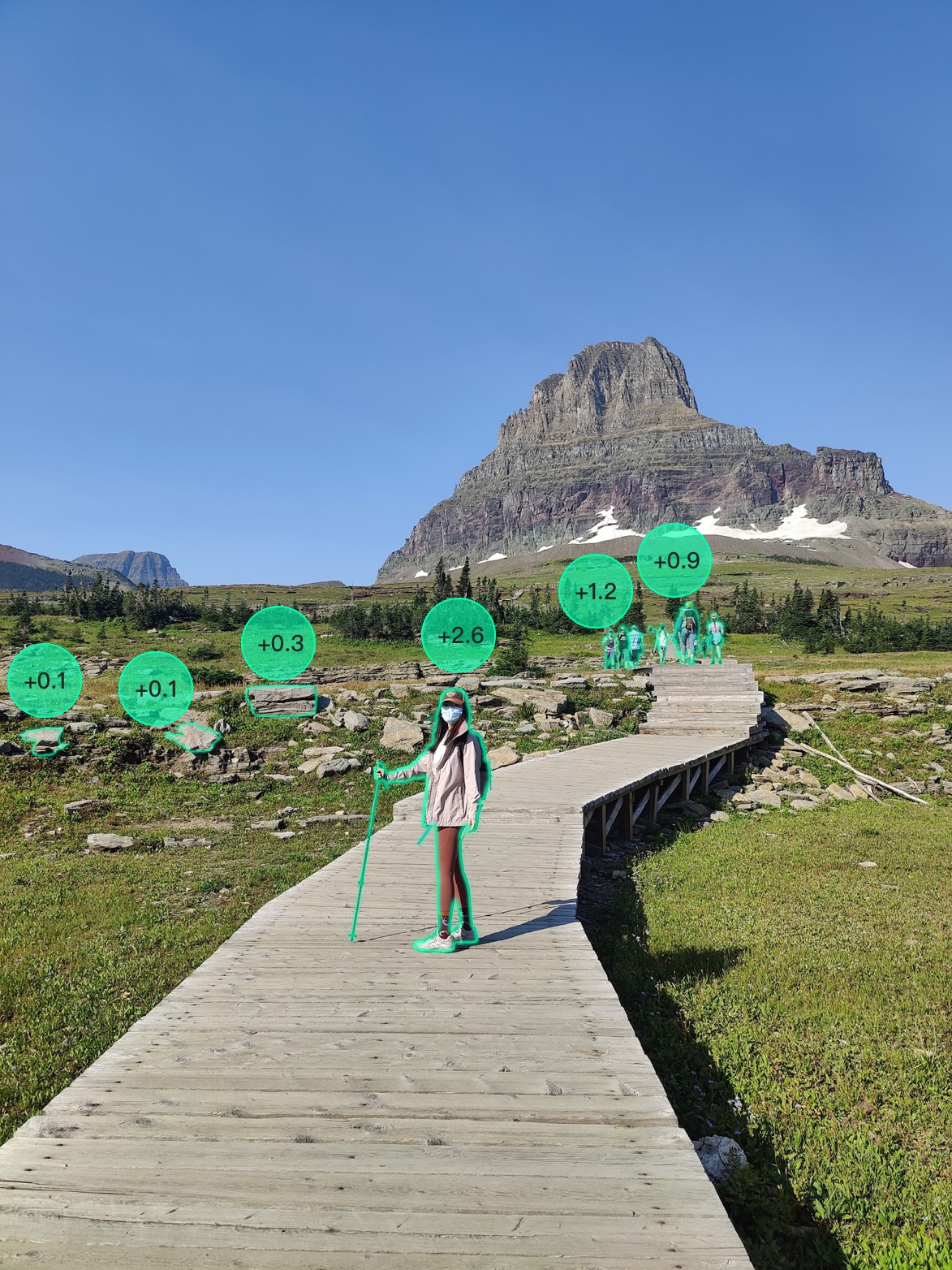}}
    \caption{Three examples of the contribution estimation results. (a) $P18$ found great results where the estimations are proportional to the size of the objects. (b) $P11$ found our algorithm is applicable to elements (stain) that do not manifest as objects. (c) $P24$'s photo where all objects contribute positively to the overall photo quality.}
    \label{fig:rank}
\end{figure}
\subsubsection{Contribution estimation of cluttered objects}
In this section, we compare the aesthetic contributions of elements estimated by our algorithm and participants. Since the rating scales may vary with different participants, we focused on the rank of the contribution values rather than their magnitude. To be specific, for each pair of elements, we checked whether the order given by our algorithm aligns with the order given by the participant. In all 108 test cases (each of the 36 photos was reviewed by 3 participants), there are a total of 14861 pairs of elements, among which 1473 ($9.91\%$) were given a different 
order. We further show the distribution of these disagreement cases at a fine-grained level in Fig.~\ref{fig:quantitative_1} (b). For each test case, we calculated the percentage of element pairs where the algorithm and the participant reviewer disagrees. We can see that most test cases ($70.74\%$) have a disagreement degree of less than $30\%$.

The low degree of conflict suggests that participants generally agree with the predictions made by our algorithm. $P18$ supported that our system provides reasonable estimations: "\emph{In the photo of the avenue, there are four vehicles. The bigger the vehicle is, the lower the contribution value. I like these results. They are intuitively right. The bigger vehicle is indeed more distracting.}" (Fig.~\ref{fig:rank} (a))

$P11$ was also satisfied with the contribution estimations (Fig.~\ref{fig:rank} (b)): "\emph{I agree the stains are intrusive. I ignored them because I was in a hurry as the speedboat passed quickly. Besides, the main subject and the buildings are indeed visually appealing.}" In $P11$'s picture, we can see that our algorithm is applicable to elements in the form of both objects and non-object. Nevertheless, $P11$ did not agree with the relative contribution of the two buildings, and she thought that we missed other buildings in our system. Despite these shortcomings, she held that "\emph{basically, I find the results really interesting and helpful}". $P24$ held a similar positive opinion on our algorithm upon seeing the results (Fig.~\ref{fig:rank} (c)).

When rating the overall accuracy of our clutter detection model, participant reviewers agreed that our system provides a good classification that aligns well with their aesthetics (Mdn = 6, IQR = 6-7).

\begin{figure}
    \centering
    \subfigure[Photo taken by $P9$.]{\includegraphics[height=0.23\linewidth]{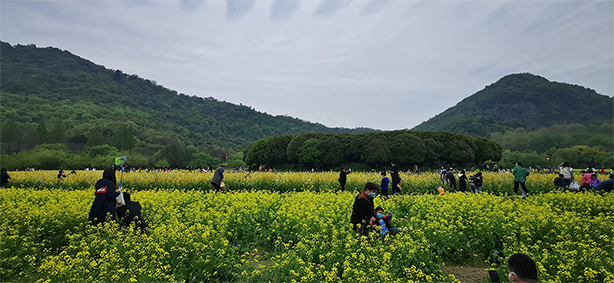}}\hfill
    \subfigure[Photo taken by $P2$.]{\includegraphics[height=0.23\linewidth]{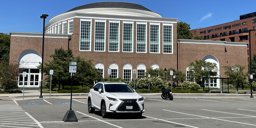}} \\
    \subfigure[$P9$'s photo after clutter removal.]{\includegraphics[height=0.23\linewidth]{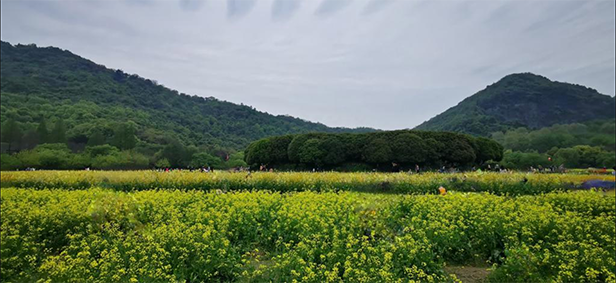}}\hfill
    \subfigure[$P2$'s photo after clutter removal.]{\includegraphics[height=0.23\linewidth]{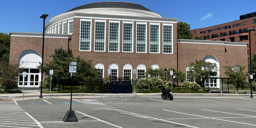}} \\
    \subfigure[Removal result of the baseline algorithm~\cite{guo2021image}.]{\includegraphics[height=0.23\linewidth]{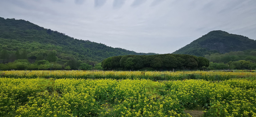}}\hfill
    \subfigure[Removal result of the baseline algorithm~\cite{guo2021image}.]{\includegraphics[height=0.23\linewidth]{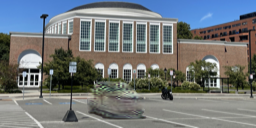}}\\
    \caption{Results of our clutter removal algorithm. (a,c,e) $P9$ found the algorithm got a superior result where the missing regions originally occupied by tourists were filled by the plants' textures in high quality. The baseline algorithm~\cite{guo2021image} cannot deal with high-resolution images, and the result is of lower clarity. (b,d,f) In the case of $P2$, the baseline contains many artifacts in the inpainted area.}
    \label{fig:removal}
\end{figure}

\subsubsection{Clutter Removal}

The answers to the Likert questions reveal that participants found the inpainted images generally contain visually plausible and semantically reasonable content. In Fig.~\ref{fig:quantitative_1} (c), we compare our clutter removal algorithm against the baseline~\cite{guo2021image}. We can see that our algorithm is of higher visual, semantic, and overall quality. We hypothesize that the baseline did not perform well because it (and most state-of-the-art image inpainting algorithms) is not specially designed for high-resolution image inpainting. We also provide two examples comparing the image inpainting results in Fig.~\ref{fig:removal}. Moreover, our algorithm has the advantage of time efficiency (Fig.~\ref{fig:quantitative_1} (c)). For each image, our algorithm takes 107±6ms while the baseline method takes 287±26ms (avg±std of 108 test cases on a 2080Ti GPU are shown).

Most participants were satisfied with our image inpainting algorithm: "\emph{I like the photo after removing other tourists. To be honest, the result is far beyond my expectation. Even when I check the photo carefully, I can hardly distinguish the generated contents from real images.}" ($P9$) In this case (Fig.~\ref{fig:removal} (a, c)), the missing regions of other tourists are filled with realistic textures of plants in the background. 

Other participants found that the capture-time clutter removal function saves their time and helps them explore more design ideas during shooting. For example, $P2$ (Fig.~\ref{fig:removal} (b, d)) told us: "\emph{The clutter removal button helped me see the scene without the vehicle with high fidelity. I was planning to remove it during the post-processing stage using tools like PhotoShop. However, after removing the vehicle, I found that the composition of the photo became unbalanced. Capture-time clutter removal helped me find this problem in advance, and I was able to solve it on the spot and explore more ideas.}" A similar experience was also reported by 11 other participants, which can partially explain why the image inpainting algorithm was triggered 2.7 times on average for the staging of one scene.

\begin{figure}
    \centering
    \includegraphics[width=\linewidth]{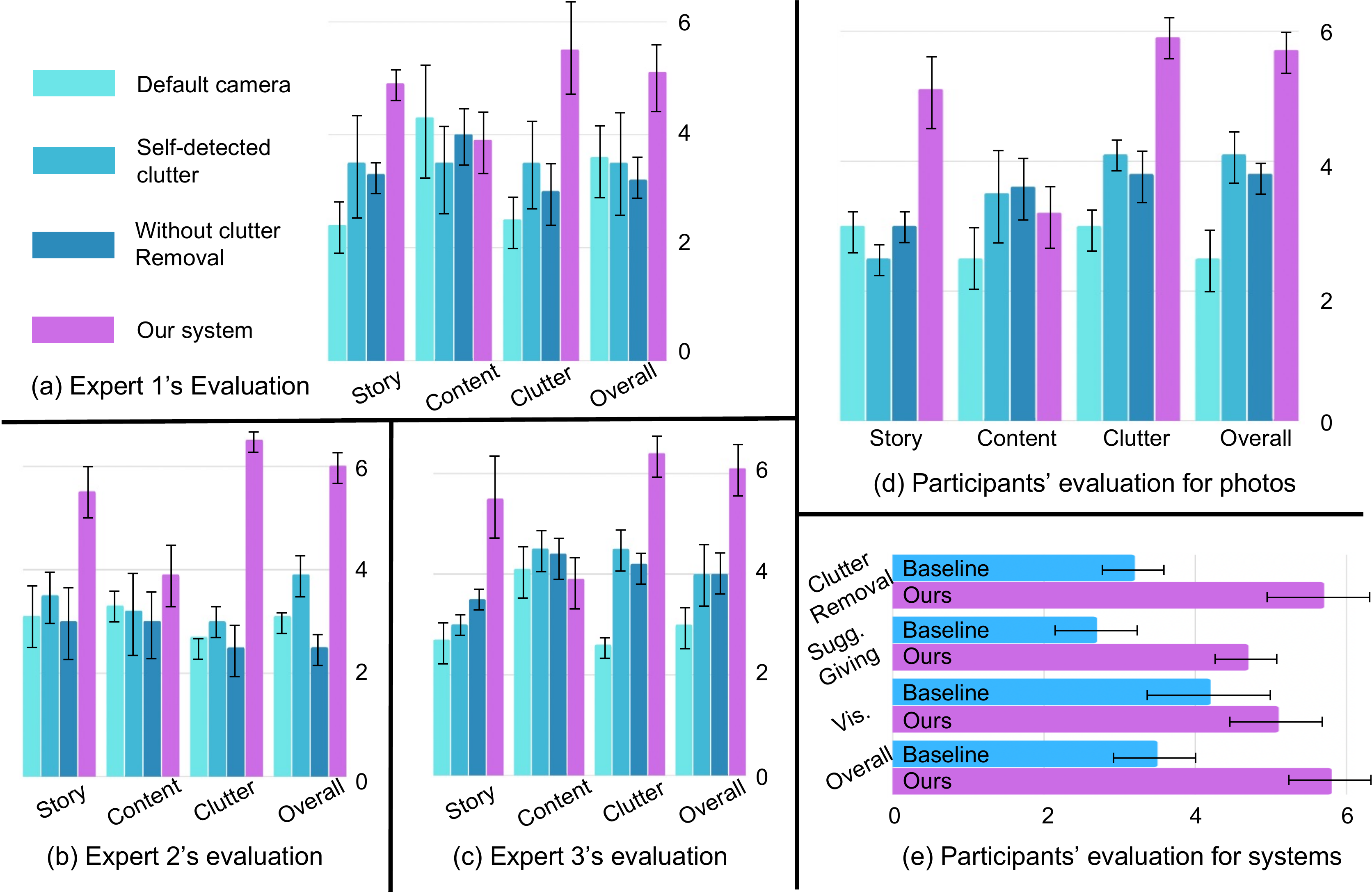}
    \caption{Experts and peer evaluations for our system against three baseline systems. (a-c) Expert evaluations of photos taken using different systems. (d) Participant evaluations of photos taken using different systems. (e) Participant evaluations regarding the interaction, visualization, and overall guidance of our system.}
    \label{fig:quantitative_2}
\end{figure}

\subsection{System Test Results}
As stated in Sec.~\ref{sec:exp-setup}, participants were tasked to take photos using three baseline systems and our systems. Their photos were then evaluated by other experimenters and three photography experts. Average ratings with standard deviations are shown in Fig.~\ref{fig:quantitative_2} (a-d). As we can see, our system has a significant advantage over all baseline systems of improving the photos in terms of story-telling, clutter reduction, and the overall quality. For example, Expert 1 thought "\emph{the proposed system can largely reduce the influence of clutter}". Expert 1 gave an average rating of 5.3 to our system's clutter reduction performance, while other systems were rated 2.2-3.1. We observed a strong correlation between clutter reduction and storytelling. All experts and 25 (out of 32) participants mentioned that clutter exerts a significant influence on the delivery of intended stories.

When ablating the clutter detection module from our system, while the photographers generally thought they had excluded clutter elements that interfere with story-telling, observers still frequently found themselves being distracted. For example, other experimenters rated 2.3 for the story-telling aspect for photos taken under this baseline system. This observation shows that photography beginners are often unaware of the clutter in their pictures. When ablating the clutter removal module from our system, 17 (out of 32) photographers reported that they encountered clutter that was not easily to be excluded, and 13 (out of 32) photographers told us they explored fewer design ideas because of less feedback during shooting. Their experience was reflected on the ratings: this baseline system got unsatisfactory scores in both experts' and participants' evaluations.

As for the reasons why the participants prefer our system, some users \textbf{noticed the clutter that they ignored}. For example, $P3$ realized that the vehicles and pedestrians were unrelated to the photo's theme (Fig.~\ref{fig:detection} (b)) and excluded them after being reminded by our system. Similarly, $P11$ was familiar with her environment and included the stain in the scene unconsciously, which lowered the quality of her photo (Fig.~\ref{fig:rank} (b)). Our system found this problem and reminded her of it.

In addition, participants thought our system \textbf{provides intuitive guidance on framing and compositions}. For example, $P19$ was not sure whether the pillars should be present (Fig.~\ref{fig:detection} (a)). Our system convinced her that the plumbing contributes negatively to the overall quality of the photo. $P19$ thus made up her mind to exclude the pillars.

In addition, participants found \textbf{the image inpainting function gives opportunities to take photos in cluttered environments.} For example, $P9$ said: "\emph{Without the clean button, I might give up this scene} (Fig.~\ref{fig:removal} (a)). \emph{There was an endless stream of tourists, and I could hardly get a clean photo.}" $P2$ held a similar opinion because, in his photo (Fig.~\ref{fig:removal} (b)), the vehicle occludes the main subject but is immovable: "\emph{Normally, I would change a scene. However, things are different with this system... That is amazing.}"

\begin{figure}
    \centering
\includegraphics[width=\linewidth]{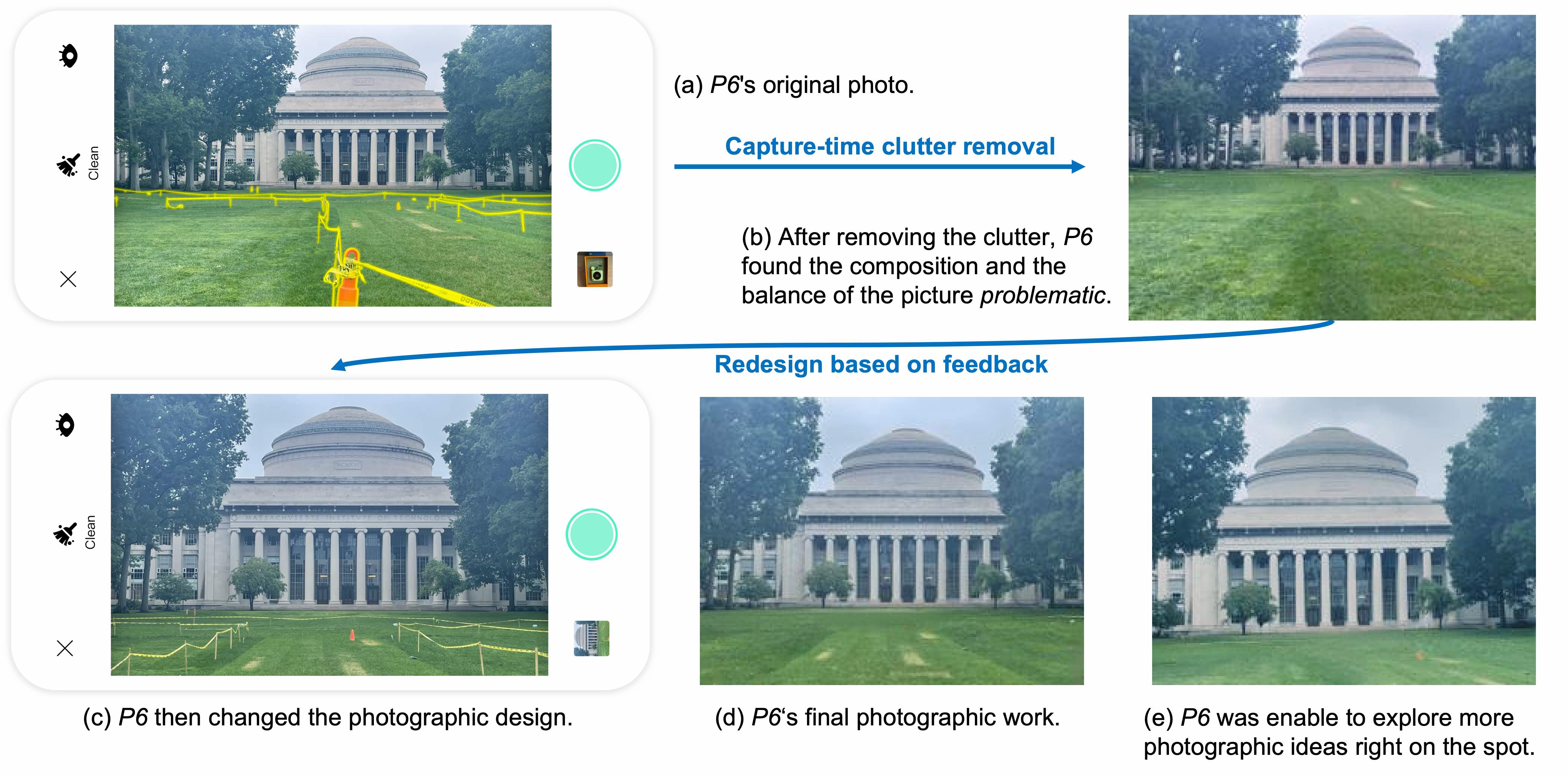}
    \caption{An example ($P6$) where incorporating clutter detection and removal in capture-time interaction encourages photographic design. If clutter handling is deferred to the post-processing stage, the user needs to return to the spot for fixing the composition problem in (b). With our system, the user was able to get more pictures with higher quality (d, e) right on the spot.}
    \label{fig:pipeline}
\end{figure}

Some participants held that \textbf{the real-time clutter removal function helps them explore more.} For example, $P6$ (Fig.~\ref{fig:pipeline}) said: \emph{"After removing the cordon ropes, I found the composition changed and the color of the picture became less vivid. Without the real-time inpainting function, this problem could only be pinpointed during post-processing."} Fixing this issue, $P6$ was able to take better photos (Fig.~\ref{fig:pipeline} (d, e)) on the spot. As $P6$ said: "\emph{this function freed me from returning to the spot to fix the composition and color problems.}" Capture-time feedback enabled him to quickly test many photography ideas on the spot. 


Other participants found they \textbf{were motivated to be bolder in their photographic work}. For example, $P28$ said: "\emph{In my daily life, I like taking photos of dishes and desserts. I am always concerned that my photo contains much clutter when there are too many items. I think the system did a good job on my photo, so I am more encouraged to capture photos of dishes with the system's guidance. I believe the system can give good suggestions and guarantee that I do not make big mistakes.}"


Overall, as shown in Fig.~\ref{fig:quantitative_2} (e), when compared to baseline systems, participants found our interactions user-friendly, covering many aspects of clutter avoidance, and can help improve the photo quality. For example, $P7$ thought that our system provides various functions using a few interactions: "\emph{When the system provider introduced how to use, I thought the interactions are more than what I need. However, during shooting, every time I expected the system to have a function, it never let me down. I think the functions cover a large number of user needs with respect to clutter handling.}"

\section{Conclusion}

In this paper, we introduce an in-camera capture-time guidance system that helps photographers identify and handle visual clutter in photos. This system is motivated by the observation that frequently appearing visual clutter often spoils the intended story of the photographer. We carry out a survey and classify clutter into three categories. We then develop a computational model to estimate the contribution of objects to the overall quality of photos, based on which we provide an interface that allows interactive identification of clutter. Suggestions and tools for dealing with different clutter are proposed. In particular, we develop an image inpainting algorithm to computationally cope with the clutter that is not cumbersome to remove. Users find the interactions provided by our system flexible, useful, and covering multiple aspects of clutter handling. For future work, we believe a generative model based on the current system that not only removes clutter but also proposes semantically reasonable and visually appealing image compositions is an exciting research direction.

\bibliographystyle{ACM-Reference-Format}
\bibliography{sample-base}

\appendix

\section{Training Scheme}\label{sec:training_scheme}

We train our clutter classification model on the public image aesthetic dataset AADB~\cite{kong2016photo}. This dataset includes 10,000 images, and, for each image, we use its overall aesthetic quality score and its attribute score evaluating whether the content is interesting. These two scores are provided by five different human raters. Original images presented in this dataset are of different sizes. To fit our model, we resize the images to $256\times 256\times 3$. 

We set the loss weight $\lambda_{\text{aes}}$ to $1$ to pay equal attention to aesthetics and content in the model. The model is trained for $100$ epochs with mini-batches of size $32$. Optimization is carried out by an Adam optimizer~\cite{kingma2014adam} with a learning rate of $4\times 10^{-4}$. An early stop mechanism is adopted to avoid model overfitting, and we stop training when the total loss does not decrease in $15$ consecutive epochs. We clip gradients with a norm larger than 5 to avoid dramatic network changes. A pre-trained ResNet with 101 layers is used for feature extraction. We do not use the fully-connected layers of the pre-trained ResNet model because the downstream task is different. The extracted feature maps are of the size $14\times 14\times 2048$. The extracted image features are also fed into the mixing network for generating the weights of the linear transformations. The mixing network is a two-layer fully-connected network with ReLU activation. The hidden layer has $128$ neurons.

The image inpainting model is trained on the ImageNet~\cite{deng2009imagenet} and Places2~\cite{zhou2017places} dataset. Corrupted images are the input to the generator, and the original images in the dataset are used as ground truth for calculating the reconstruction loss and the discriminator loss. Similar to~\citet{zeng2020high}, we corrupt the images with two kinds of masks. The first is object masks, where we remove objects detected by Mask R-CNN from images. The second is random strokes to avoid a bias towards missing regions in the shape of objects. The generator consists of an encoder and a decoder. The encoder is a 6-layer convolutional neural network (CNN) with $\{48,48,96,96,192,192\}$ kernels at each layer. The decoder is also a CNN but has 7 convolutional layers with $\{192,192,96,96,48,$ $24,3\}$ kernels. An Adam optimizer with a learning rate of $1\times 10^{-4}$ is used to train mini-batches that contain 64 training samples.

\end{document}